\documentclass[aps,showpacs, nofootinbib,floatfix]{revtex4}
\usepackage{graphicx, graphics, color}

\input amssym.tex
\input amssym.def

\newcommand{\be}{\begin{equation}}
\newcommand{\ee}{\end{equation}}
\newcommand{\ben}{\begin{eqnarray}}
\newcommand{\een}{\end{eqnarray}}

\newcommand{\la}{{\lambda}}

\newcommand{\cO}{{\cal O}}

\newcommand{\p}{\partial}
\newcommand{\na}{\nabla}

\begin{document}

\title{Analytic study on backreacting holographic superconductors with dark matter sector}

\author{{\L}ukasz Nakonieczny and Marek Rogatko}
\email{rogat@kft.umcs.lublin.pl, 
marek.rogatko@poczta.umcs.lublin.pl,
lnakonieczny@kft.umcs.lublin.pl}
\affiliation{Institute of Physics \protect \\
Maria Curie-Sklodowska University \protect \\
20-031 Lublin, pl.~Marii Curie-Sklodowskiej 1, Poland }


\date{\today}
\pacs{11.25.Tq, 04.50.-h, 98.80.Cq}

\begin{abstract}
The variational method for Sturm-Liouville eigenvalue problem was employed to study
analytically properties of the holographic superconductor with dark matter sector, in which
a coupling between Maxwell field and another $U(1)$-gauge field was considered.
The backreation of the dark matter sector on gravitational background in question was also
examined.

\end{abstract}

\maketitle
\section{Introduction}
The AdS/CFT correspondence \cite{mal} 
plays a significant role in understanding the strongly coupled field theories.
It relates string theory on asymptotically anti-de Sitter spacetime (AdS)
to a conformal field theory on the boundary and enables to establish method for obtaining
correlation functions in a strongly interacting field theory by means of a dual classical gravity
description \cite{wit}. 
The aforementioned correspondence provides a theoretical insight into description of Hall
and Nernst effects \cite{har07,har07a} (see also some
contemporary reviews and references therein \cite{reviews}). 
Moreover, it has been also successful applied to 
nuclear physics problems like heavy ion collisions at the Relativistic Heavy Ion Collider (RHIC) \cite{mat07}.
\par
It has been also proposed that AdS/CFT correspondence could be useful in
describing superconductor phase transitions, namely when an Abelian symmetry is broken outside an 
AdS black hole this phenomenon results in the formation 
of the holographic superconductor (formation of charged scalar condensation) in the dual CFT \cite{gub05}-\cite{gub08}.
It was indicated that the expectation value of charged operators underwent the $U(1)$-symmetry breaking
and the second order phase transition took place \cite{har08}-\cite{har08b}. 
Other properties of superconducting materials bounded with 
the influence of the magnetic field, were also reconstructed. The aforementioned studies covered the topic 
of critical magnetic 
fields and Abrikosov vortices \cite{prop}-\cite{vortices}. At the beginning, studies of the 
holographic superconductivity
when no backreations of matter fields were considered. Contrary to the previously elaborated
cases, the backreaction caused that even uncharged scalar fields could form a condensation in $(2+1)$-dimensions
\cite{har08b}. In the case of p-wave holographic models it was shown that the phase transition leading to the formation of 
vector hair, changed from the second order to the first one, depending on the strength of the gravitational coupling 
\cite{amm10}-\cite{cai11}.
These facts have triggered studies of the holographic superconductivity including the backreaction effect of 
the matter fields in question (see, e.g., Refs.\cite{gub09}-\cite{bri10}). Recently, also the backreaction between 
holographic insulator and superconductor
was considered \cite{hor10}-\cite{bri11}.   

\par
The other theories of gravity were also taken into account. Holographic superconductivity in 
Einstein-Gauss-Bonnet (EGB) gravity were elaborated and among all it was revealed that
the higher curvature corrections made the condensations harder to occur
\cite{gre09}-\cite{kan11}. 
Realization of holographic superconductivity in Einstein-Maxwell-dilaton gravity being
a generalized model built from the most extensive covariant gravity Lagrangian with at most two
derivatives of fields \cite{apr10} was given in Ref.\cite{liu10}. On the other hand,
gravity theory including dilaton field were also investigated
in the case of the potential models of superfluids and superconductors \cite{sal12}.
Horava-Lifshitz gravity \cite{horava}
and string/M theory \cite{mth} were also paid attention to. Additionally other kind of electro-magnetic
theories were studied due to the holographic superconductivity (see for example \cite{bi} and references therein) .
For instance, the presence of the Born-Infeld scale parameter decreases the critical temperature
and the ratio of the gap frequency in conductivity, comparing to the Maxwell electrodynamics.
Recently, also frameworks of non-linear electrodynamics acquire much attention due to holographic superconductivity
studies \cite{nonlin}. On the other hand, analytical 
methods of investigations of the holographic superconductivity
with or without a backreaction were examined in Refs.\cite{anal} (see e.g., for the selected examples of this still
growing literature).
\par
Furthermore, it naturally arises a question about possible matter configurations in 
AdS spacetime. In Ref.\cite{shi12}
it was revealed that strictly stationary Einstein-Maxwell
spacetime with negative cosmological constant
could not allow for the existence of nontrivial
configurations of complex scalar fields or form fields. This statement was also confirmed
in the case of strictly stationary, simply connected Einstein-Maxwell-axion-dilaton spacetime with negative cosmological
constant and arbitrary number of $U(1)$-gauge fields \cite{bak13}. 
\par
Present observations of the Universe envisage the fact that
it may be dominated by an {\it exotic} kind of fields 
never observed directly. In the late 1990s the independent surveys for distant
supernovae type Ia (Sn Ia), calibrated as {\it standard candles},  
announced that the high-redshifted supernovae of this type appeared about $40$ percent fainter or
equivalently more distant than expected in a flat, matter-dominated Universe \cite{accuniv}.
Other investigations like analysis of cosmic microwave background \cite{ben03} or baryonic 
acoustic oscillations \cite{eis05} confirm that our Universe is filled with the {\it exotic} matter
with negative pressure, causing its acceleration.
\par
The contemporary observations provide also the strong evidence that almost 22 percent of the total energy
density of our Universe is of a form of dark matter, which on its turn is unclear what it is made of.
Recently, the new model mimicking the dark matter sector was proposed (see e.g., for the latest issues Refs.\cite{model}).
In the aforementioned prediction the standard model is coupled to the dark sector
via an interactive term. It has the form of a direct coupling between $U(1)$-gauge strength tensor of
Maxwell field and the dark matter one. 
It worth pointing out that, the interaction of $U(1)$ Abelian-Higgs field model having cosmic string
solutions with a low energy dark matter sector being also a $U(1)$ Abelian-Higgs model was studied
in the context of cosmic string interacting with dark string \cite{dark matter}.

In order to generalize existing results concerning holographic superconductors as well as
motivated by the above astrophysical observations, we shall study
the influence of the proposed dark matter sector coupled to the standard model on holographic
superconductivity. We shall consider bottom up theory and study backreaction of the considered 
matter fields on gravitational background in question.

\par
The paper is organized as follows. In Sec.II we derive equations of motion 
for a holographic superconductor model with backreactions in n-dimensional AdS black hole background, in the
underlying theory. Sec.III is devoted to the analytical description of superconducting phase,
while in Sec.IV the critical temperature dependence on the spacetime dimension as well as parameters
characterizing dark matter sector are examined. In Sec.V we present remarks and conclusions of our investigations.

\section{Action and equations of motion}
In this section one considers a holographic Abelian-Higgs model for s-wave superconductor. In the matter
sector we assume a coupling between Maxwell gauge field and another $U(1)$-gauge field in order to represent
the dark matter influence. We remark that two gauge uncoupled fields were studied in the case of
the so-called {\it unbalanced holographic superconductor} \cite{mus}. On the other hand, 
holographic Abelian-Higgs model for s-wave superconductivity with a coupling between the gauge field and
a new massive gauge field introduced in order to describe impurities was examined in Refs.\cite{imp}.
In what follows we shall for the first time study backreaction of dark matter sector on the gravitational background.
Perhaps considering the phase transition temperature enables one to find justification and evidence of
the aforementioned dark matter model in the future experiments.
\par
To commence with we shall specify the gravitational part of the action which implies
\ben
S_{g} = \int \sqrt{-g}~ d^n x~  \frac{1}{2 \kappa^2}\bigg( R - 2\Lambda\bigg), 
\een
where $\kappa^2 = 8 \pi G_{n}$ is an n-dimensional gravitational constant.
The cosmological constant will be given by
$\Lambda = - \frac{(n-1)(n-2)}{2L^2}$, 
where $L$ is the radius of the AdS spacetime.
The underlying matter sector is modelled by the Abelian-Higgs sector coupled to the second $U(1)$-gauge field.
It is provided by the following action:
\be
\label{s_matter}
S_{m} = \int \sqrt{-g}~ d^nx  \bigg( 
- \frac{1}{4}F_{\mu \nu} F^{\mu \nu} - \left [ \nabla_{\mu} \psi - 
i q A_{\mu} \psi \right ]^{\dagger} \left [ \nabla^{\mu} \psi - i q A^{\mu} \psi  \right ]
- V(\psi) - \frac{1}{4} B_{\mu \nu} B^{\mu \nu} - \frac{\alpha}{4} F_{\mu \nu} B^{\mu \nu}
\bigg), 
\ee  
where the scalar field potential satisfies
$V(\psi) = m^2 |\psi|^2 + \frac{\lambda_{\psi}}{4} |\psi|^4$.
$F_{\mu \nu} = 2 \nabla_{[ \mu} A_{\nu ]}$
stands for the ordinary Maxwell field strength tensor, while
the second $U(1)$-gauge field $B_{\mu \nu}$ is given by  
$B_{\mu \nu} = 2 \nabla_{[ \mu} B_{\nu ]}$. Moreover, $m,~ \lambda_{\psi},~ q$ represent
mass, a coupling constant and charge related to the scalar field $\psi$, respectively.
On the other hand, $\alpha$ is a coupling constant for both $U(1)$ fields.
In order to be compatible with the current observations it should be on the order of $10^{-3}$.
\par
The motivation standing behind our studies is related to the fact that one has the strong
evidence that almost 22 percent of the total energy density of our Universe is in the form of dark
matter. Up till now we have only vague idea what it is made of. The model in which dark matter sector
is coupled to the Standard Model has its possible roots in astrophysical observations of $511$ keV gamma
rays observed by Integral/SPI \cite{integral} and experiments revealing electron and positron excess observed
in galaxy by ATIC \cite{atic} and PAMELA \cite{pamela}. It turned out that the energy of the aforementioned
excess is in the range between a few of GeV to a few of TeV. The conceivable explanation is the annihilation
of dark matter into electrons, i.e., below GeV scale, the interaction term can be in the form of a direct
coupling between $U(1)$-gauge strength tensor of the dark matter sector and $U(1)$ Maxwell strength tensor.
\par
There are also other possible evidences revealing new physics, e.g., the $3.6-\sigma$ discrepancy
between measured value of the muon anomalous magnetic moment and its prediction in the Standard Model \cite{muon}.
From the cosmological point of view, these kinds of dark matter models admitted cosmic string like solutions,
topological defects which might arise in the early Universe \cite{dark matter}.
In the above context it will be interesting to look for the imprints of dark matter sector in properties
of holographic superconductors and reveal the other {\it experimental} confirmation of the underlying model.

The Einstein equations for the system in question may be written in the form as
\be
R_{\mu \nu} = \bigg( \frac{n}{n-2} - 1 \bigg)\Lambda g_{\mu \nu} + 
2 \kappa^2 \bigg( T_{\mu \nu} - \frac{1}{n-2} T \bigg),
\ee
where the energy-momentum tensor for 
the matter fields yields
\ben 
T^{\mu}_{\nu} &=& F^{\mu}_{\alpha}F_{\nu}^{\alpha} + 
B^{\mu}_{\alpha}B_{\nu}^{\alpha} + \alpha B^{\mu}_{\alpha}F_{\nu}^{\alpha} 
+ 2 \nabla^{\mu} \psi \nabla_{\nu} \psi + 2 q^2 A^{\mu}A_{\nu} \psi^2 + \nonumber \\
&+& g^{\mu}_{\nu} \bigg( 
- \frac{1}{4}F_{\alpha \beta}F^{\alpha \beta} 
- \frac{1}{4}B_{\alpha \beta}B^{\alpha \beta} -\frac{\alpha}{4}F_{\alpha \beta}B^{\alpha \beta}
- \nabla_{\alpha} \psi \nabla^{\alpha} \psi - q^2 A_{\alpha} A^{\alpha} \psi^2 - V(\psi) 
\bigg).
\een
Variations of the underlying action with respect to the matter fields reveal the
exact form of the equations of motion for the considered theory. They have the forms as follows:
\ben
\label{matter_1}
\nabla_{\mu} F^{\mu \nu} &-& 2 q^2 A^{\nu} \psi^2 + \frac{\alpha}{2} \nabla_{\mu} B^{\mu \nu} = 0, \\
\label{matter_2}
\na_\mu \na^\mu \psi &-& q^2 A_{\mu} A^{\mu} \psi - \frac{1}{2} \frac{ \partial V(\psi) }{ \partial \psi} = 0, \\
\label{matter_3}
\nabla_{\mu} B^{\mu \nu} &+& \frac{\alpha}{2} \nabla_{\mu} F^{\mu \nu} = 0.
\een
Having in mind the form of the relation
(\ref{matter_3}) we see that the dynamics of the $B_{\mu}$ field is determined (up to the integration constant)
by the Maxwell field potential $A_{\mu}$.
After inserting equation (\ref{matter_3}) into the relation (\ref{matter_1}) we arrive at the following:
\be
\bigg( 1 - \frac{\alpha^2}{4} \bigg) \nabla_{\mu} F^{\mu \nu} - q^2 \psi^2 A^{\nu} = 0.
\ee 
In order to proceed further 
one introduces a general line element which enables us to consider
backreation of matter fields on the gravitational sector. Namely, it is provided by 
\be
\label{metric}
ds^2 = - f(r) e^{- \chi(r)}dt^2 + \frac{dr^2}{f(r)} + r^2 h_{i j} dx^i dx^j,
\ee 
where $f$ and $\chi$ are functions of $r$-coordinate, $h_{i j}$ is the metric tensor on the 
$(n-2)$-dimensional submanifold.
Further, we assume that the considered gauge fields posses only the temporal components which also 
depend only on the radial coordinate $r$. They can be written as
\be
A_{t} = \phi(r), \qquad B_{t} = \eta(r).
\ee
Consequently it leads to the equations of motion as follows:
\ben
\partial^{2}_{r} \psi &+& 
\bigg( \frac{n-2}{r} + \frac{1}{f} \partial_{r}f - \frac{1}{2} \partial_{r} \chi  \bigg) \partial_{r} \psi +
\frac{ e^{\chi} q^2 \phi^2}{f^2} \psi - \frac{1}{2 f} \frac{ \partial V}{ \partial \psi} = 0, \\
\partial^{2}_{r} \phi &+& \bigg( \frac{1}{2} \partial_{r} \chi + \frac{n-2}{r} \bigg) \partial_{r} \phi -
\frac{2 q^2  \psi^2}{ \tilde{\alpha} f } \phi = 0, \\
\partial_{r} \eta &=& -c_{1} r^{ - n +2} e^{- \frac{1}{2} \chi} - \frac{\alpha}{2} \partial_{r} \phi,
\een 
where we have introduced $\tilde{\alpha} = 1 - \frac{\alpha^2}{4}$ and the integration constant $c_{1}$. 
\par
On the other hand, the Ricci curvature tensor components yield
\ben
R^{t}_{t} &-& R^{r}_{r} = \frac{n-2}{2 r} f \partial_{r} \chi, \\
R^{x}_{x} &=& - \frac{1}{r} \partial_{r}f + \frac{1}{2 r} f \partial_{r} \chi - \frac{n-3}{r^2} f.
\een
One assumes that $(n-2)$-dimensional submanifold is flat and $x$ in $R^{x}_{x}$ represent 
the generic coordinate in this submanifold.
\par
By virtue of the above relations one can readily verify that the complete set of the differential 
equations describing the system in question is determined by
\ben
\label{sys_1}
\partial_{r} \chi &=& - \frac{4 \kappa^2 r}{ n -2} \bigg(  2 (\partial_{r}\psi)^2 + 
2 \frac{e^{\chi} q^2 \phi^2 \psi^2}{f^2} \bigg), \\
\label{sys_2}
\partial_{r}f &+& \frac{n-3}{r}f + \frac{2r}{n-2} \Lambda = 
- \frac{2 \kappa^2 r}{n - 2} \bigg[
2 f (\partial_{r} \psi)^2 + 2 \frac{ e^{\chi} q^2 \phi^2 \psi^2}{f} \\ \nonumber
&+& 2 V(\psi) + 
e^{\chi} \bigg(  (\partial_{r} \phi)^2 + (\partial_{r} \eta)^2 + \alpha \partial_{r} \phi \partial_{r}\eta \bigg)
\bigg], \\
\label{sys_3}
\partial^{2}_{r} \psi &+& 
\bigg( \frac{n-2}{r} + \frac{1}{f} \partial_{r}f - \frac{1}{2} \partial_{r} \chi  \bigg) \partial_{r} \psi +
\frac{ e^{\chi} q^2 \phi^2}{f^2} \psi - \frac{1}{2 f} \frac{ \partial V}{ \partial \psi} = 0, \\
\label{sys_4}
\partial^{2}_{r} \phi &+& \bigg( \frac{1}{2} \partial_{r} \chi + \frac{n-2}{r} \bigg) \partial_{r}\phi -
\frac{2 q^2  \psi^2}{ \tilde{\alpha} f } \phi = 0, \\
\label{sys_5}
\partial_{r} \eta &=& -c_{1} r^{ - n +2} e^{- \frac{1}{2} \chi} - \frac{\alpha}{2} \partial_{r} \phi.
\een  
Using equation (\ref{sys_5}) one can rewrite relation (\ref{sys_2}).
Namely, $\eta$ field can be eliminated from the set of the differential equations system,
and the modified relation (\ref{sys_2}) may be cast in the form as
\be
\partial_{r}f + \frac{n-3}{r}f + \frac{2r}{n-2} \Lambda = 
- \frac{2 \kappa^2 r}{n - 2} \bigg[
2 f (\partial_{r} \psi)^2 + 2 \frac{ e^{\chi} q^2 \phi^2 \psi^2}{f} + 2 V(\psi) + 
e^{\chi} \bigg(  \tilde{\alpha} (\partial_{r} \phi)^2  + a^2 \bigg)
\bigg],
\ee 
where we set $a^2 = c_{1}^2~ r^{-2n +4}~ e^{-\chi}$.
\par
In the normal phase one has that $\psi(r) = 0$, which in turn implies that $\chi$ 
is constant. Having this in mind the remaining equations may be rewritten in the forms as
\ben
\partial_{r}f &+& \frac{n-3}{r}f + \frac{2r}{n-2} \Lambda = 
- \frac{2 \kappa^2 r}{n - 2} e^{\chi}~ \bigg(  \tilde{\alpha} (\partial_{r} \phi)^2  + a^2 \bigg), \\
\partial^{2}_{r} \phi &+&  \frac{n-2}{r} \partial_{r} \phi  = 0.
\label{pp}
\een
It is easy to verify that the solution of (\ref{pp}) yields
\be
\phi = \mu - \frac{\rho}{r^{n-3}}, 
\ee
where $\mu$ and $\rho$ are interpreted as the chemical potential and charge density in the dual
field theory. On the other hand, for $f(r)$ one arrives at the expression
\be
f = \frac{r^2}{L^2} + \frac{2 \kappa^2}{n-2} 
\left [ \tilde{\alpha} (n-3) \rho^2 + \frac{c_{1}^2}{n-3} \right ] 
\frac{1}{ r^{2n-6}} - 
\left [ \frac{r_{+}^{n-1}}{L^2} + \frac{2 \kappa^2}{n-2} \frac{ 
\tilde{\alpha} (n-3) \rho^2 + \frac{c_{1}^2}{ n-3} }{ r_{+}^{n-3}} \right ] \frac{1}{r^{n-3}}.
\ee
It can be observed that in the case when $\psi = 0$ 
the line element under consideration represents the Reissner-Nordstr{\"o}m-AdS (RN-AdS) black hole.
\par
In the next step we shall tackle the problem when
$\psi \neq 0$. Namely, we pay attention to the appropriate boundary conditions. From the previous 
relations one has that the functions $\chi$ and $f$ satisfy at the black hole event horizon $r_{+}$
the following relations:
\ben
(\partial_{r} \chi)_{| r_{+}} &=& 
- \frac{4 \kappa^2 r_{+}}{ n -2} \left [  2 (\partial_{r}\psi)_{| r_{+}}^2 + 
2 \frac{e^{\chi(r_{+})} q^2 (\partial_{r} \phi)_{| r_{+}}^2 \psi(r_{+})^2}{(\partial_{r}f)_{| r_{+}}^2} \right ], \\
(\partial_{r}f)_{| r_{+}} &=& 
-\frac{2r_{+}}{n-2} \Lambda - \frac{2 \kappa^2 r_{+}}{n - 2} \left [
2 V(\psi(r_{+})) + 
e^{\chi(r_{+})} \left ( \tilde{\alpha} (\partial_{r} \phi)_{| r_{+}}^2 + a(r_{+})^2 \right )
\right ].
\een
On the other hand, at the black hole event horizon, one
finds that the boundary conditions yield
\ben
\phi(r_{+}) &=& 0, \\ 
(\partial_{r} f)_{| r_{+}} (\partial_{r} \psi)_{| r_{+}} &-& \frac{1}{2} 
(\frac{ \partial V}{\partial \psi})_{| r_{+}} = 0, \\
(\partial_{r} \eta)_{| r_{+}} &=& - c_{1} r_{+}^{-n +2} e^{- \chi(r_{+})} 
- \frac{\alpha}{2} (\partial_{r} \phi)_{| r_{+}} .
\een
Next, it remains to check the behaviour of the above functions
at the AdS boundary (when $r \rightarrow \infty$). 
Their asymptotical forms are provided by the following:
\be
\chi \rightarrow 0, \quad f \sim \frac{r^2}{L^2}, \quad \phi \sim \mu - \frac{\rho}{r^{n-3}}, \quad 
\psi \sim \frac{\psi_{-}}{r^{\Delta_{-}}} + \frac{ \psi_{+}}{r^{\Delta_{+}}}, \quad
\partial_{r} \eta  = - c_{1} r^{-n +2}  - \frac{\alpha}{2} \partial_{r} \phi ,
\ee  
where the exponent is defined by the relation
$\Delta_{\pm} = \frac{1}{2}[ n - 1 \pm \sqrt{ (n-1)^2 + 4m^2} ]$.

\section{Analytical description of superconducting phase}
In this section we shall implement Sturm-Liouville method \cite{pol02} to the analysis of
behaviour of s-wave holographic superconductor phase transition with backreaction on the gravitational
sector of the underlying theory. To proceed further, one introduces new variable $z = \frac{r_{+}}{r}$
and rewrite the equations of motion in the forms as  
\ben \label{aa}
\partial_{z} f &-& \frac{n-3}{z} f + \frac{ (n-1)r_{+}^2}{L^2 z^3} -
\frac{1}{z^3} \frac{2 \kappa^2 r_{+}^2}{n-2} \bigg( 
2 f \bigg( \frac{z^4}{r_{+}^2} (\partial_{z} \psi)^2 + \frac{ e^{\chi} q^2 \phi^2 \psi^2}{f^2}  \bigg) 
\\ \nonumber
&+& 2 V(\psi) + e^{\chi} \bigg( \tilde{\alpha} \frac{z^4}{r_{+}^2} (\partial_{z} \phi)^2 + 
\frac{c_{1}^2}{r_{+}^{2n - 4}} z^{2n-4}   \bigg) \bigg) = 0,
\\ \label{bb}
\partial_{z} \chi &-& \frac{1}{z^3} \frac{4 \kappa^2 r_{+}^2}{n-2} \bigg( 
\frac{2 z^4}{r_{+}^2} (\partial_{z} \psi)^2 + 2 \frac{ e^{\chi} q^2 \phi^2 \psi^2}{f^2}
\bigg) = 0, \\ \label{cc}
\partial^{2}_{z} \phi &-& \bigg(  \frac{n-4}{z} - \frac{1}{2} \partial_{z} \chi \bigg) \partial_{z} \phi 
+ \frac{2 r_{+}^2}{ \tilde{\alpha} z^4} \frac{q^2 \psi^2}{f}\phi = 0, \\
\partial^{2}_{z} \psi &-& \bigg(  \frac{n-4}{z} + \frac{1}{2} \partial_{z} \chi - \frac{1}{f} 
\partial_{z}f  \bigg) \partial_{z} \psi 
+ \frac{r_{+}^2 e^{\chi} q^2 \phi^2}{f^2 z^4} \psi
- \frac{r_{+}^2}{z^4} \frac{1}{2f} \frac{ \partial V(\psi)}{\partial \psi}  =0,\\ \label{dd}
\partial_{z} \eta &=& \frac{ c_{1} e^{- \frac{1}{2} \chi} }{r_{+}^{n-3}} z^{n-4} - \frac{ \alpha}{2} \partial_{z} \phi.
\een
In order to
gain insight into the influence of the scalar filed on the background geometry, we solve the above 
system of differential equations perturbatively by expanding
the unknown functions in the series in some small parameter $\varepsilon$. 
Because of the fact that the value of the scalar operator $< O_i >$, where $i = -,~+$, is small
in the vicinity of the critical point, one introduces it as an expansion parameter
$\varepsilon = < O_i >$ and expands the functions in question as follows:
\ben
f &=& f_{0} + \varepsilon^2 f_{2} +  \varepsilon^4 f_{4} + \dots ,\\
\chi &=& \varepsilon^2 \chi_{2} + \varepsilon^4 \chi_{4} + \dots,\\
\phi &=& \phi_{0} + \varepsilon^2 \phi_{2} + \varepsilon^4 \phi_{4} + \dots, \\
\psi &=& \varepsilon \psi_{1} + \varepsilon^3 \psi_{3} + \dots, \\
\eta &=& \eta_{0} + \varepsilon^2 \eta_{2} + \varepsilon^4 \eta_{4} + \dots.
\een
On this account, the chemical potential $\mu$ can be written in the form \cite{her10}
\be
\mu = \mu_{0} + \varepsilon^2 \delta \mu,
\ee 
where $\delta \mu > 0$. It reveals that near the phase transition the order parameter
may be described as a function of the chemical potential, i.e., 
$\varepsilon = ((\mu - \mu_{0})/ \delta \mu)^{1/2}$. The phase transition can take place when
$\mu \rightarrow \mu_0$.
Consequently, it can be observed that $\varepsilon$ approaches to zero and the critical value of the chemical
potential is given by $\mu_c = \mu_0$.\\
At the zeroth order, 
the metric function will be provided by RN-AdS black hole spacetime, described in the preceding
section. The new radial variable $z$ enables us to find the exact form of $\phi_{0}$ and $f_{0}$. 
Having in mind expressions in the nearby of the critical point and using equation (\ref{cc}),
one concludes that
\be
\phi_{0} = \mu_{0}( 1 - z^{n-3}),
\ee
where $\mu_0 = \rho/{r_+}^{n-3}$. On the other hand, at critical point when
$\mu = \mu_0 = \rho/{r_{+c}}^{n-3}$, where $r_{+c}$ is the value of black hole event horizon radius 
at the critical point, we obtain
\be
\phi_{0} = \lambda r_{+c} (1 - z^{n-3}),
\label{ll}
\ee
with $\la = \rho/r_{+c}^{n-2}$. The value $r_{+c}$, being
value of the black hole event horizon at critical temperature, is bounded with the temperature 
in the sense that Hawking's black hole temperature
depends on the value of the event horizon radius. It is provided by the relation
\be
T_{BH} = {1 \over 4 \pi}~\p_r g_{tt} \mid_{r=r_{event~ hor}}.
\ee
Just the temperature of a black hole with event horizon radius value equal to $r_{+c}$
represents the critical temperature above which the condensation 
cannot take place. \\
Consequently, taking into account equation (\ref{aa}) and the relation (\ref{ll})
one achieves that $f_0$ is given by
\be
f_{0} = r_{+}^2 \left [ 
\frac{1}{L^2 z^2} - \frac{z^{n-3}}{L^2} - z^{n-3}(1 - z^{n-3}) 
\frac{2 \kappa^2}{n-2} \left ( \frac{\tilde{\alpha} (n-3) \lambda^2 r_{+c}^2}{r_{+}^2} 
+ \frac{ c_{1}^2}{ (n-3) r_{+}^{2n-4}}  \right )
\right ],
\ee 
The lowest order corrections are solely determined by the $\psi$ field. The equation for 
this field (up to ${\cal O}(\varepsilon^2)$ terms) implies
\be
\partial^{2}_{z} \psi_{1} - \bigg(  \frac{n-4}{z} - \frac{1}{f_{0}} \partial_{z} f_{0}  \bigg)~\partial_{z} \psi_{1} 
- \bigg(  \frac{ m^2 r_{+}^2}{f_{0} z^4} - \frac{ r_{+}^2 q^2 \phi_{0}^2 }{z^4 f_{0}^2} \bigg) \psi_{1} = 0,
\ee
where one uses the fact that 
$\frac{\partial V(\psi)}{ \partial \psi} \approx 2 m^2 \varepsilon \psi_{1} + {\cal O}(\varepsilon^3)$. 
On the other hand, the asymptotic behaviour of the scalar field near the AdS boundary 
is provided by $\psi_{1} = \psi_{i} \frac{ z^{\Delta_{i}}}{r_{+}^{\Delta_{i}}}$,
where $i = \pm$ and $\Delta_{i} = \frac{1}{2} \left[ (n-1) \pm \sqrt{ (n-1)^2 + 4m^2}  \right]$, while $\psi_{i}$
represents the expectation value of the adequate operator on the CFT side. 
Namely we choose as $\psi_{i}$ the expansion parameter $ < O_{i}>$. As the
temperature leads to the critical one, scalar field $\psi$ approaches the limit which is
even valid at $T=T_c$. Next, in order to match the behaviour at the boundary one defines
$\psi_{1}$ in the form as follows:
\be
\psi_{1} = \frac{ <O_{i}>}{r_{+}^{\Delta_{i}}} z^{\Delta_{i}}F(z),
\ee
where we introduce a trial function $F(z)$ normalized to the following value $F(0) = 1$.
By virtue of the above relation we arrive at the equation
\be
\label{eq_F_org}
\partial^{2}_{z}F + \left [ \frac{2 (\Delta_{i} +2) -n}{z} + \frac{\partial_{z} g}{g} \right ] \partial_{z} F
+ \left [ \frac{\Delta_{i}(\Delta_{i} - n + 3)}{z^2}  + \frac{\Delta_{i}}{z} \frac{\partial_{z}g}{g} - \frac{m^2}{ z^4 g} 
+ \frac{q^2 \lambda^2 r_{+c}^2 (1 - z^{n-3})^2}{r_{+}^2 z^4 g^2}  \right ]F = 0.
\label{st1}
\ee
To simplify the above equation we introduce a function $g$ related to $f_0$ by the following:
\be
f_0 = r_{+}^2 \left \{  
\frac{1}{L^2 z^2} - \frac{z^{n-3}}{L^2} - z^{n-3}(1 - z^{n-3}) 
\left [ \frac{2 \kappa^2}{n-2} \left (  \frac{ \tilde{\alpha}(n-3) 
\lambda^2 r_{+c}^2}{r_{+}^2} + \frac{c_{1}^2}{(n-3) r_{+}^{2n-4}} \right )   \right ]
\right \} = r_{+}^2 g.
\ee
In the next step we shall rewrite equation (\ref{st1}) in the form Sturm-Liouville form.
Namely it can be done if one defines
\be
T = z^{2 \Delta_{i} + 1} \left \{ 
(n-2)(z^{1 - n} - 1) - L^2 2\kappa^2 \left [  
\frac{ \tilde{\alpha}(n-3) \lambda^2 r_{+c}^2}{r_{+}^2} + \frac{c_{1}^2}{(n-3) r_{+}^{2n-4}}  \right ] 
(1 - z^{n-3})
\right \},
\ee
and after a little of algebra we find that we have left with 
\be
\label{eq_F_SL}
\partial_{z} ( T \partial_{z} F) + 
T \left \{ \frac{\Delta_{i}}{z} \left [  \frac{\Delta_{i} + 3 - n}{z} + 
\frac{ \partial_{z} g}{g} \right ] - \frac{ m^2}{z^4 g} 
+ \lambda^2 \frac{ r_{+c}^2 q^2 (1 - z^{n-3})^2}{r_{+}^2 z^4 g^2} \right \} F = 0.
\ee
The above equation is solved subject to the boundary condition $\p_z F(0) = 0$.

\section{Critical temperature}
Now we refine our studies to the problem of finding the critical temperature below which the
condensation of $\psi$ field can occur. An important notion bounded with the phase transition
is the temperature connected with black hole in question. 
Using the definition of the Hawking temperature which for the line element in question 
\be
T_{BH} = {1 \over 4 \pi}~\p_r f(r)~e^{\chi(2) \over 2}~\mid_{r = r_+},
\ee
one can verify that, up to $\cO(\varepsilon)$,
it is provided by the relation
\be
T_{BH} = \frac{r_{+}}{4 \pi} \left \{ 
\frac{n-1}{L^2} - \frac{2 \kappa^2}{n-2} (n-3) 
\left [ \frac{\tilde{\alpha} (n-3) \lambda^2 r_{+c}^2}{r_{+}^2} 
+ \frac{c_{1}^2}{(n-3) r_{+}^{2n-4}}  \right ]
\right \},
\ee
where $c_{1}$ is a constant, which can be partially restricted by the demand that 
we do not introduce any new parameters on the CFT side and the fact that 
the critical temperature scales with charge density in a typical way. 
In view of these postulates
we calculate the black hole temperature at $r_{+} = r_{+c}$. It yields
\be
\label{Tc_BH}
T^{c}_{BH} = \frac{1}{4 \pi} \left \{ 
\frac{n-1}{L^2} - (n-3) \frac{2 \kappa^2}{n-2} \left [ (n-3) \tilde{\alpha} \lambda^2 + \tilde{c}^2 \right ] 
\right \} \left ( \frac{\rho}{\lambda} \right )^{\frac{1}{n-2}},
\ee
where we use the relation $c_{1}^2 = \tilde{c}^2 (n-3) r_{+}^{2n -4}$.\\
On this account we shall find the critical temperature. 
Namely, having in mind 
equations (\ref{eq_F_SL}) and (\ref{Tc_BH}),
multiplying both sides of the relation (\ref{eq_F_SL}) by the trial function $F(z)$ and integrating
over the interval under consideration, we can specify
$\lambda^2$ as a spectral parameter of the Sturm-Liouville eigenvalue problem \cite{pol02}.
On the other hand, the expression for 
estimating the minimum eigenvalue of 
$\lambda^2$ is provided by
\be
\label{lambda}
\lambda^2 \leq \frac{ \int \limits_{0}^{1} dz  \left [ T~ (\partial_{z} {F})^2   
- T~ U~ {F}^2 \right ] }{ \int \limits_{0}^{1} dz~ T~V ~{F}^2  },
\ee
where the quantities $U$ and $V$ are defined by
\ben
U &=& \frac{\Delta_{i}}{z} \left [  \frac{\Delta_{i} + 3 - n}{z} + \frac{ \partial_{z} g}{g} \right ] 
- \frac{ m^2}{z^4 g}, \\
V &=& \frac{ r_{+c}^2 q^2 (1 - z^{n-3})^2}{r_{+}^2 z^4 g^2}.
\een
The validity of the use of the Sturm-Liouville method demands that our trial function 
must fulfill one of the prescribed sets of the boundary conditions \cite{pol02}. 
Meanwhile beside this formal boundary conditions our test function should also 
be in the agreement with the physical boundary condition ($F(0) = 1,~ 
\partial_{z} F(0) = 0$). Having this in mind we choose the following test function    
$F = 1 - az^2 + \frac{2}{3} a z^3$, where $a$ is a free parameter.
The considered trial function satisfies all the aforementioned requirements. 
Integration of equation (\ref{lambda}) 
enables us to obtain $\lambda^2$ as a function of a free parameter $a$ \cite{anal}.
Further, one finds the minimum of this function and use it in the calculations of the critical 
temperature, the gravitational constant $\kappa$ is a backreacting parameter. 
\par
Having in mind  equation (\ref{Tc_BH}), we remark that in comparison to the 
single $U(1)$ field case one has two new parameters being integration constants $\tilde{c}$ and the 
$\tilde{\alpha} = 1 - \frac{\alpha^2}{4}$, where
$\alpha$ is a coupling constant between two gauge fields in the theory. 
Black hole temperature behaviour depends on whether 
$\tilde{c}^2$ is equal to $|(n-3) \tilde{\alpha} \lambda^2|$ or not.\\
Let us analyze the first case 
$\tilde{c}^2 \neq |(n-3) \tilde{\alpha} \lambda^2|$. In this 
situation the exact value of $\tilde{c}$
is unimportant from the point of view of the qualitative behavior of the black hole temperature. 
The sign of the relation $(n-3) \tilde{\alpha} \lambda^2 + \tilde{c}^2$ is important because of the 
$\alpha$-coupling dependence. Namely, if $\tilde{c} = 0$ then for $\alpha < 2$ it 
reaches a positive value, while $\alpha > 2$ it is negative. 
Taking into account $\tilde{c} \neq 0$, one has that
we need to consider the larger value of $\alpha$ to obtain a negative sign in the aforementioned expression. 

\begin{table}[h!]
\begin{center}
\begin{tabular}{|l|l|l|}
  \hline
  \multicolumn{3}{|c|}{$n = 5$, $\Delta = 3$, $m^2 = -3$, $L = 1$, $q = 1$} \\
  \hline
  $\kappa$ & $\alpha = 0.5$ ($\tilde{\alpha} = 0.94$)& $\alpha = 2.5$ ($\tilde{\alpha} = -0.5625$) \\
  \hline
  $\kappa = 0.0$ & $T^{c}_{BH} = 0.1977 \rho^{\frac{1}{3}}$ & $T^{c}_{BH} = 0.1977 \rho^{\frac{1}{3}}$ \\
  \hline 
  $\kappa = 0.05$ & $T^{c}_{BH} = 0.1926 \rho^{\frac{1}{3}}$ & $T^{c}_{BH} = 0.2007 \rho^{\frac{1}{3}}$ \\
  \hline
  $\kappa = 0.1$ & $T^{c}_{BH} = 0.1774 \rho^{\frac{1}{3}}$ & $T^{c}_{BH} = 0.2099 \rho^{\frac{1}{3}}$ \\
  \hline
  $\kappa = 0.15$ & $T^{c}_{BH} = 0.1533 \rho^{\frac{1}{3}}$ & $T^{c}_{BH} = 0.2256 \rho^{\frac{1}{3}}$ \\
  \hline
  $\kappa = 0.2$ & $T^{c}_{BH} = 0.1223 \rho^{\frac{1}{3}}$ & $T^{c}_{BH} = 0.2484 \rho^{\frac{1}{3}}$ \\
  \hline
  $\kappa = 0.25$ & $T^{c}_{BH} = 0.0882 \rho^{\frac{1}{3}}$ & $T^{c}_{BH} = 0.2795 \rho^{\frac{1}{3}}$ \\
  \hline
  $\kappa = 0.3$ & $T^{c}_{BH} = 0.0558 \rho^{\frac{1}{3}}$ & $T^{c}_{BH} = 0.3201 \rho^{\frac{1}{3}}$ \\
  \hline
\end{tabular}
\end{center}
\caption{The behavior of the critical temperature on the backreacting parameter $\kappa$ in the case of five-dimensional black hole, $\tilde{c} = 0$.}
\label{tab1}
\end{table}

\begin{table}[h!]
\begin{center}
\begin{tabular}{|l|l|l|}
  \hline
  \multicolumn{3}{|c|}{$n = 4$, $\Delta = 2$, $m^2 = -2$, $L = 1$, $q = 1$} \\
  \hline
  $\kappa$ & $\alpha = 0.5$ ($\tilde{\alpha} = 0.94$)& $\alpha = 2.5$ ($\tilde{\alpha} = -0.5625$) \\
  \hline
  $\kappa = 0.0$ & $T^{c}_{BH} = 0.1182 \rho^{\frac{1}{2}}$ & $T^{c}_{BH} = 0.1182 \rho^{\frac{1}{2}}$ \\
  \hline
  $\kappa = 0.05$ & $T^{c}_{BH} = 0.1169 \rho^{\frac{1}{3}}$ & $T^{c}_{BH} = 0.1190 \rho^{\frac{1}{3}}$ \\
  \hline
  $\kappa = 0.1$ & $T^{c}_{BH} = 0.1128 \rho^{\frac{1}{2}}$ & $T^{c}_{BH} = 0.1215 \rho^{\frac{1}{2}}$ \\
  \hline
  $\kappa = 0.15$ & $T^{c}_{BH} = 0.1062 \rho^{\frac{1}{3}}$ & $T^{c}_{BH} = 0.1256 \rho^{\frac{1}{3}}$ \\
  \hline
  $\kappa = 0.2$ & $T^{c}_{BH} = 0.0974 \rho^{\frac{1}{2}}$ & $T^{c}_{BH} = 0.1315 \rho^{\frac{1}{2}}$ \\
  \hline
  $\kappa = 0.25$ & $T^{c}_{BH} = 0.0870 \rho^{\frac{1}{3}}$ & $T^{c}_{BH} = 0.1394 \rho^{\frac{1}{3}}$ \\
  \hline
  $\kappa = 0.3$ & $T^{c}_{BH} = 0.0756 \rho^{\frac{1}{2}}$ & $T^{c}_{BH} = 0.1496 \rho^{\frac{1}{2}}$ \\
  \hline
\end{tabular}
\end{center}
\caption{The behavior of the critical temperature on the backreacting parameter $\kappa$ in the case of 
four-dimensional black hole, $\tilde{c} = 0$.}
\label{tab2}
\end{table}

In Table \ref{tab1} we see that for a positive $\tilde{\alpha}$ and
increasing value of the backreacting parameter $\kappa$,
one obtains decreasing critical temperature. On the contrary, for the negative $\tilde{\alpha}$
the situation is quite opposite.
For the completeness of reasoning we present also results for four-dimensional 
black hole in the Table \ref{tab2}. 
One can draw a conclusion that the change of the dimension of the underlying
spacetime does not qualitatively change the phase transition temperature dependence on
$\kappa$ and $\alpha$.
To conclude, we may state that in the case $\tilde{c}^2 \neq |(n-3) \tilde{\alpha} \lambda^2|$ 
one receives the two
different types of behaviours of $T^{c}_{BH}(\kappa)$, i.e., for 
$(n-3) \tilde{\alpha} \lambda^2 + \tilde{c}^2 > 0$ the temperature decreases 
with the increase of $\kappa$ and for 
the case when $(n-3) \tilde{\alpha} \lambda^2 + \tilde{c}^2 < 0$,
the temperature in question increases as $\kappa$ increases.
\par
On the other hand, for $\tilde{c}^2 = |(n-3) \tilde{\alpha} \lambda^2|$ we have 
only one type of behavior of $T^{c}_{BH}(\kappa)$, i.e., 
for the positive value of $\tilde{\alpha}$ when $\kappa$ increases,
the temperature in question decreases. For a negative $\tilde{\alpha}$,
the critical temperature is independent of $\kappa$ and equal to the $\kappa = 0.0$ case.
In Figs.1 and 2 we depicted the aforementioned behaviour, i.e.,
the dependence of the critical temperature on $\kappa$ for four and five-dimensional spacetimes,
having in mind the positive and negative values of the coupling constant $\tilde{\alpha}$. 
It should be also stressed that in the case of four-dimensional spacetime, when
$\kappa$ is equal to zero our results are in a perfect agreement with the ones received in Ref.\cite{har08}.
\par
Summing it all up, it is worth mentioning that
the behaviour responsible for the negative value of $\tilde{\alpha}$ is rather surprising, because
of the fact that the condensed phase appears at high temperature, not at low as usual.
We refer this action as a {\it retrograde condensation} \cite{ret}. The similar behaviour
was observed in Ref.\cite{apr11}, where holographic superconductors were obtained from gauged
supergravity theory with a complex scalar field.

\section{Conclusions}
In our paper we have used the variational method for the Sturm-Liouville eigenvalue problem for analytical
studies of the holographic superconductor including backreactions of matter field sector on
gravitational background.
In the matter sector we assume a coupling between Maxwell gauge field and another
$U(1)$-gauge field in order to mimic the possible dark matter sector influence on the critical temperature.
One examined the cases of five and four-dimensional gravitational background as well as
two possibilities of choosing the coupling constant $\tilde{\alpha}$. In the case when
$\tilde{\alpha}> 0$ one obtains that the bigger backreacting parameter we choose the smaller
critical temperature we get. This behaviour was true for both five and four-dimensional 
holographic superconductors. On the contrary, when we take into account
$\tilde{\alpha} < 0$, the obtained results recall the {\it retrograde condensation phenomenon} \cite{ret}.
Namely,
the bigger $\kappa$ we imply the bigger critical temperature one achieves.
The {\it retrograde condensation} was also observed in the case of gauged supergravity
holographic superconductors \cite{apr11}. In the limit of $\kappa = 0$, when we have no backreaction,
our analytical method gave the same value of the critical temperature as received by numerical methods in
Ref.\cite{har08}, i.e., $T^{c}_{BH} = 0.1182 \rho^{\frac{1}{2}}$.
\par
For the future researches, it will be interesting to examine the influence of the magnetic field on
the dark matter sector holographic superconductor and the influence of the other gauge fields on superconductor 
parameters.
We hope to prepend these problems elsewhere.


\begin{acknowledgments}
{\L}N was supported by the Polish National Science Centre under doctoral scholarship number 2013/08/T/ST2/00122.   
MR was partially supported by the grant of the National Science Center
$2011/01/B/ST2/00488$.
\end{acknowledgments}


\begin{figure}[H]
\label{fig1}
\includegraphics[scale=0.35,angle= -90]{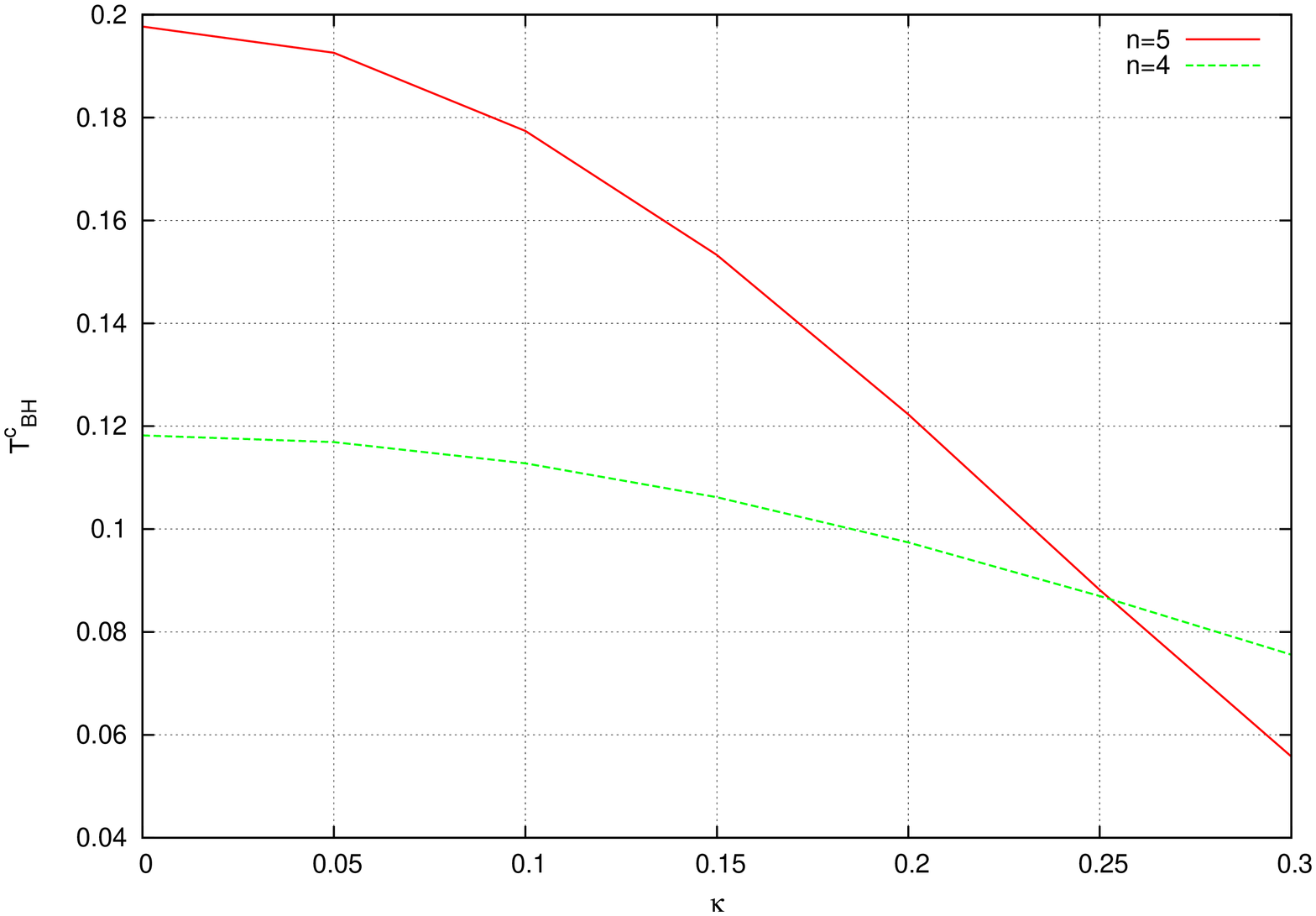}
\caption{(color online) The critical temperature as a function of $\kappa$ for the two different dimensions of spacetime. 
The parameters are chosen as $\tilde{\alpha} = 0.94$, $L = 1$, $q = 1$, and  $\Delta =3$, $m^2 = -3$ for $n=5$
while $\Delta =2$, $m^2 = -2$ for $n=4$.}
\end{figure}

\begin{figure}[H]
\label{fig2}
\includegraphics[scale=0.35,angle= -90]{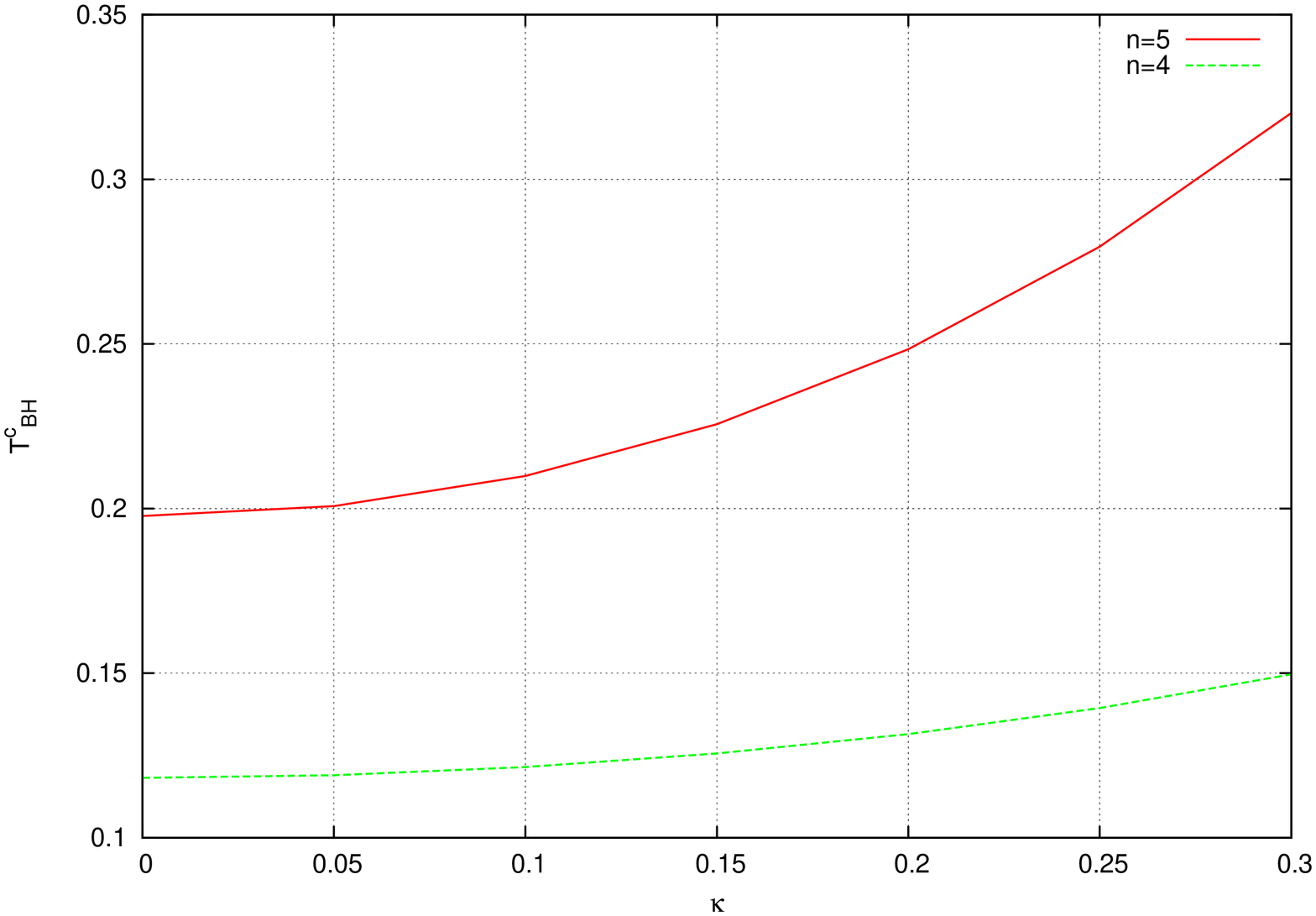}
\caption{(color online) The critical temperature as a function of $\kappa$ for two different dimensions of spacetime. 
The parameters are set as $\tilde{\alpha} = -0.5625$, $L = 1$, $q = 1$, and  $\Delta =3$, $m^2 = -3$ for $n=5$
while $\Delta =2$, $m^2 = -2$ for $n=4$.}
\end{figure}

\end{document}